\documentclass[twocolumn,showpacs,preprintnumbers,amsmath,amssymb]{revtex4}
\usepackage{graphicx}
\usepackage{dcolumn}
\usepackage{bm}

\newcommand{\ie}{{\it i.e.}}

\def\ave#1{\langle~#1~\rangle}

\begin{document}

\title{Strain-assisted spin manipulating and the discerption of
strain-induced spin splitting}

\author{Yuan Li and You-Quan Li}
\affiliation{Zhejiang Institute of Modern Physics and Department of
Physics,\\ Zhejiang University, Hangzhou 310027, P. R. China}

\begin{abstract}
We show that the efficiency of manipulating electron spin
in semiconductor quantum wells can be enhanced by tuning the
strain strength. The effect combining intrinsic and
strain-induced spin splitting varies
for different systems, which provides an alternative route to
understand the experimental phenomena brought in by the strain.
The types of spin splittings caused by strain are suggested to be
distinguished by the measurement of the
electron-dipole-spin-resonance intensity through changing the
direction of the $ac$ electric field in the $x$-$y$ plane of the
quantum well and tuning the strain strengths.

\end{abstract}

\received{\today} 

\maketitle

\section{Introduction}

Manipulating electron spins by an external electric field is a
central issue in the realization of spintronics on the basis of
solid-state
materials~\cite{Wolf,Datta,Schliemann,Zutic} as it
is important for quantum computing and information
processing~\cite{Loss}. In the experiment by Kato et
al.~\cite{Kato1} electron spins are manipulated by means of the
voltage-controlled g-tensor modulation technique which is applicable
for materials with small g-tensor merely. Recently, a mechanism
called electric dipole spin resonance (EDSR) was proposed to
investigate spin manipulating for electrons in a parabolic quantum
well~\cite{Rashba,Efros}. An in-plane or a perpendicular
electric field was shown to efficiently manipulate the spin for
electrons in quantum wells. In that mechanism a tilted magnetic
field is required but small g-tensor is not necessary.

It has been recognized recently that a strain-induced spin
splitting is able to effectively manipulate electron spins without
magnetic fields, which provides an alternative route, called
strain engineering, for solid-state spin manipulation. For
example, it has been employed to control electron-spin
precession~\cite{Kato2,Crooker1,Beck} in
zinc-blende structure semiconductors and to tune the spin
coherence with a significant enhancement of the spin dephasing
time~\cite{Wu}. In semiconductor epilayers, the effect of the
strain on electron-spin transport~\cite{hur} and that of uniaxial
tensile strain on spin coherence~\cite{Knotz,Sih}  have also been
carried out in recent experiments. The types of strain-induced
spin splittings in strained bulk semiconductors were
analyzed~\cite{Zhang} theoretically, whereas, it is not very clear
so far that which type of spin splitting (i.e., Rashba-type or
Dresselhuas-type) produced by the strain~\cite{Kato2,Zhang} plays
an important role in manipulating electron spins. Since the spin
manipulating in terms of strain is easily realizable in practical
devices, it will be useful to exhibit the type of strain-induced
spin splitting.

In this paper, we show that the efficiency of spin manipulation for
electrons can be enhanced through adjusting the strain strength of
the semiconductor quantum wells. The effect combining intrinsic and
strain-induced spin splitting varies for different systems. We
propose a method to identify whether the spin splitting induced by
strain is of Dresselhuas-type or Rashba-type. This paper is
organized as follows: In next section, we give a general
consideration on the electron-dipole spin resonance caused by both
the intrinsic and strain-induced spin splitting. In
Sec.~\ref{sec:Dresselhaus} and Sec.~\ref{sec:Rashba}, we consider
the systems with intrinsic spin-orbit coupling of Dresselhaus type
and Rashba-type respectively. The spin manipulation for an InSb
quantum well imposed with strain is investigated by means of the
orbital mechanism. As a comparison, the combining effects of
intrinsic and strain-induced spin splitting in some other kind of
semiconductor quantum wells are also studied. A summary of our main
conclusion is given in the last section.

\section{General consideration}

For electron gases in quantum wells or two-dimensional
heterostructures of certain semiconductors,
there exist two intrinsic
spin-orbit couplings called Dresselhaus and Rashba couplings that
arise from the bulk-inversion asymmetry or the
structure-inversion asymmetry of the material, respectively.
The Dresselhaus type spin-orbit interaction
can be obtained by averaging the corresponding bulk
expression over the motion relevant to the confined degree of
freedom~\cite{Dyakonov}.
If the first electron subband is merely populated
in $[001]$ quantum wells with the growth direction along the
$z$-axis,
the Dresselhaus type Hamiltonian $H_{in}^{\mathrm  D}$ is given by,
\begin{eqnarray}\label{eq:intrinsicD}
H_{in}^{\mathrm  D} =\sigma_xk_x(\lambda k_y^2-\beta)+\sigma_yk_y(\beta-\lambda
k_x^2),
\end{eqnarray}
where $\beta=\lambda\langle k_z^2\rangle$
with $\lambda$ denoting the Dresselhaus spin-orbit coupling strength
and $\langle k_z^2\rangle$ being averaged over the ground state.
Here $H_{in}^{\mathrm  D}$ contains both linear and cubic terms in $k$.

The Rashba spin-orbit interaction can be written
as~\cite{Rashba4,Bychkov},
\begin{eqnarray}\label{eq:intrinsicR}
H_{in}^{\mathrm  R} =\alpha(\sigma_xk_y-\sigma_yk_x),
\end{eqnarray}
where $\alpha$ refers to the Rashba spin-orbit coupling strength.
In some systems, the Rashba or the Dresselhaus spin-orbit coupling dominates over
the other effects. In order to manipulate electron spins
efficiently, it becomes important to know the relative strengths
between Rashba and Dresselhaus interactions in the system under
consideration.

In zinc-blende type semiconductors, strain introduces additional
spin splitting which can be of structure-inversion-asymmetry
type~\cite{Meier},
\begin{eqnarray}\label{eq:strain1}
H_{st}^{\mathrm  R}
&=&\frac{1}{2}C_3[\sigma_x(\epsilon_{xy}k_y-\epsilon_{xz}k_z)
+\sigma_y(\epsilon_{yz}k_z-\epsilon_{yx}k_x)
   \nonumber \\
&&+\sigma_z(\epsilon_{zx}k_x-\epsilon_{zy}k_y)].
\end{eqnarray}
It can also be of bulk-inversion-asymmetry type
if the diagonal elements $\epsilon_{ij}$ are included
\begin{eqnarray}\label{eq:strain2}
H_{st}^{\mathrm  D} &=&D[\sigma_xk_x(\epsilon_{zz}-\epsilon_{yy})
+\sigma_yk_y(\epsilon_{xx}-\epsilon_{zz})
   \nonumber \\
&&+\sigma_zk_z(\epsilon_{yy}-\epsilon_{xx})].
\end{eqnarray}
Here $C_3$ and $D>0$ are material constants, and $\epsilon_{ij}$
($i,j=x,y,z$) denotes the symmetric strain tensor.
We call $H_{in}^{\mathrm  D}$ and $H_{in}^{\mathrm  R}$ in Eqs.~(\ref{eq:intrinsicD}) and
(\ref{eq:intrinsicR}) intrinsic spin-orbit couplings so as to
distinguish them from the strain-induced spin splittings in
Eqs.(\ref{eq:strain1}) and (\ref{eq:strain2}).
These four types of spin-orbit interactions may take place
simultaneously if strain is exerted on a sample.
However, they can play different roles in manipulating
electron spins. It is worthwhile to study which kind of
strain-induced spin splittings plays an important role in spin
manipulation in various semiconductor quantum wells.

In order to carry out a general calculation of
electron-dipole-spin-resonance intensity
for semiconductor quantum wells with spin-orbit interactions,
we consider
\begin{eqnarray}\label{eq:hamiltonian}
H=H_0 + H_{so} + e\mathbf{E}(t)\cdot \mathbf{r},
\end{eqnarray}
which describes two-dimensional electrons with spin-orbit coupling
in a parabolic quantum well. An in-plane {\it ac} electric field
$\mathbf{E}(t)=E(t)(\cos\phi, \sin\phi,0)$ and a tilted magnetic
field $\mathbf{B}(\theta, \varphi)$ with $\theta$ and $\varphi$
being the polar and azimuthal angle of $B$ together with a strain
are applied to the system. Accordingly, the first part in
Eq.~(\ref{eq:hamiltonian}) reads
$$
H_0 = \frac{1}{2m^\star}
   (\mathbf{p} + \frac{e}{c}\mathbf{A})^2
 +\frac{1}{2}m^{\star}\omega_0^2 z^2
    + \frac{g}{2} \mu^{}_B\mathbf{\sigma}\cdot\mathbf{B},
$$
where $\mathbf{A}$ is the vector potential of the tilted magnetic
field, $m^{\star}$ denotes the effective mass of the electron,
and $\omega_0$ characterizes the parabolic potential well.
The second
part $H_{so}$ in Eq.~(\ref{eq:hamiltonian}) may include either
intrinsic or strain-induced spin-orbit couplings.

To diagonalize the Hamiltonian $H_0$, one needs to rotate the original
coordinate frame $\{\hat x, \hat y, \hat z \}$ to the new one
$\{\hat x', \hat y', \hat z' \}$,
where $\hat z'$ is chosen in alignment with the
orientation of $\mathbf{B}$, $\hat y'$ is lying in $x$-$y$ plane, and $\hat x'$
is chosen to form a right-hand triple with $\hat y'$ and $\hat z'$.
Thus the coordinates in both frame systems are related,
$(\hat{x}, \hat{y},
\hat{z})= (\hat{x}', \hat{y}', \hat{z}')R^T$, by
\begin{eqnarray*}
R= \left( {\begin{array}{*{100}c}
   \cos\theta \cos\varphi, & -\sin\varphi, & \sin\theta\cos\varphi \\[1mm]
   \cos\theta \sin\varphi, &  \cos\varphi, & \sin\theta\sin\varphi \\[1mm]
   -\sin\theta,            &      0,       & \cos\theta  \\
\end{array}} \right),
\end{eqnarray*}
which also relates the momentum components in the two frame systems,
$k_i=R_{ij}k'_{j}$ (here $i,j=x,y,z$).
By using the Landau gauge $\mathbf{A}=(0,Bx',0)$,
$H_0$ can be written as the sum of two harmonic oscillators~\cite{Efros}
and its energy levels are given by,
\begin{eqnarray}
E_s(n^{}_+, n^{}_-)= \hbar\omega_{+}\bigl(n_{+}+\frac{1}{2}\bigr)
                   + \hbar\omega_{-}\bigl(n_{-}+\frac{1}{2}\bigr)
     +\frac{s}{2}\hbar\omega_z,
     \nonumber\\
\end{eqnarray}
where $s=\pm 1$ label the spin states and $n_\pm$ refer to the
orbital quantum numbers, $\omega_\pm(\theta)$ are the frequencies of
the coupled cyclotron-confinement modes
\begin{eqnarray}
\omega_\pm^2(\theta) &=&
  \frac{\omega_0^2+\omega_c^2
  \pm \Delta^2 ~\mathrm {sgn}(\omega_0-\omega_c)}{2},
   \nonumber\\[2mm]
\Delta^2 &=&
(\omega_0^4+\omega_c^4-2\omega_0^2\omega_c^2\cos2\theta)^{1/2}.
\end{eqnarray}

If the spin-orbit coupling $H_{so}$ is relatively small in
comparison to other energy scales, such as the confinement energy
$\hbar\omega_0$, the cyclotron energy $\hbar\omega_c=\hbar
eB/m^{\star}c$ and Zeeman-splitting energy $\hbar\omega_z=g\mu_BB$,
one can calculate the EDSR intensity by employing
the method proposed in Ref.~\cite{Efros}.
The strategy of this method is to eliminate the terms related to
spin-orbit couplings in the original Hamiltonian
with the help of a canonical transformation $e^{F}$.
After some algebraic calculations, the operators of coordinates and momenta
in original coordinate can be expressed as linear combinations of
the creation and annihilation operators of the harmonic
oscillators.
Furthermore, the spin-orbit coupling $H_{so}$
and the interaction term $e \mathbf{E}(t)\cdot \mathbf{r}$
can be expressed in terms of the creation and annihilation operators.
The EDSR intensity  $I \propto\, \mid T\mid^2$
is obtained by evaluating the matrix $T$ which characterizes the
spin-flip transitions induced by the $ac$ electric field
$\mathbf{E}(t)$, namely,
\begin{eqnarray}\label{eq:EDSR}
T=\frac{1}{E(t)}\langle n_+,n_-,\uparrow \mid \mathbf{E}(t)\cdot \bigl[\hat{F},\,
\mathbf{r}\bigr]\mid n_+,n_-,\downarrow\rangle.
\end{eqnarray}
Here the operator $\hat{F}$ is perturbatively
determined by the following relation
\begin{eqnarray}\label{eq:Foperator}
&& \langle n_+',n_-',s'\mid \hat{F}\mid n_+,n_-,s \rangle
  \nonumber\\
&& = \frac{\langle n_+',n_-',s' \mid H_{so} \mid n_+,n_-,s \rangle
}{E_{s'}(n_+',n_-')-E_{s}(n_+,n_-)}
 + \textrm{high order}.
\end{eqnarray}
This is the condition for the cancellation of the spin-orbit
coupling term $H_{so}$ with the first order (or further orders if necessary)
perturbation terms
brought in by the canonical transformation $e^{\hat{F}}$ which
connects the eigenstates of $H_0 + H_{so}$ with the eigenstates of
$H_0$.
Note that the Pauli matrices as well as ${\mathbf r}$ appeared in Eq.~({\ref{eq:hamiltonian}})
are with respect to the original coordinate frame $\{\hat x, \hat y, \hat z\}$.
They should also be reexpressed with respect to the
new coordinate frame $\{\hat x', \hat y', \hat z'\}$
so that Eqs.~(\ref{eq:EDSR}) and (\ref{eq:Foperator}) are computable.

\section{Systems with intrinsic Dresselhaus coupling}\label{sec:Dresselhaus}

\subsection{Strain-induced Rashba-type coupling}

We firstly consider the strain-induced Rashba-type spin splitting
for the systems with intrinsic Dresselhaus coupling
(we call D+R case for brevity).
An concrete example of
such a system is an InSb quantum well~\cite{Liu}, where $H_{in}^{\mathrm  D}$
dominates over $H_{in}^{\mathrm  R}$ with typical value
$\lambda=200 \,\mathrm {eV\AA^3}$~\cite{Perel'}. The application of diagonal strain on the
InSb quantum well in the $x,y,z=[100],[010],[001]$ directions does
not introduce any observable spin splitting but the shear strain
leads to a splitting described by $H_{st}^{\mathrm  R}$~\cite{Zhang,Seiler}.
For compression along the $[110]$ axis,
$\epsilon_{xy}=\epsilon'_{110}/2$ and the strain
$\epsilon'_{110}$ is given by
$$\epsilon'_{110}=\epsilon_{110}-\epsilon_{1\bar{1}0}=\frac{1}{2}S_{44}P_{110},$$
where $S_{44}$ and $P_{110}$ are the compliance coefficient and the
applied stress, respectively. Accordingly,
$\epsilon_{xz}$=$\epsilon_{yz}=0$ and the electric field is in-plane
$(\ave{k_z}=0)$, then the spin precession for two dimensional electrons
in the InSb quantum well under the above strain configuration is
described by the following Hamiltonian
\begin{eqnarray}\label{eq:hso1}
H_{so}^{\mathrm  D+R} &=&H_{in}^{\mathrm  D}+H_{st}^{\mathrm  R}\nonumber\\
&=&\beta(\sigma_yk_y-\sigma_xk_x)+\gamma(\sigma_xk_y-\sigma_yk_x)\nonumber\\
&&+\lambda(\sigma_xk_xk_y^2-\sigma_yk_yk_x^2),
\end{eqnarray}
where the strain parameter $\gamma=\frac{1}{2}C_3\epsilon_{xy}$,
$C_3=1.13\times 10^{-7}\mathrm {eVcm}$~\cite{Bernevig,Ranvaud}.
The former two terms in Eq.~(\ref{eq:hso1}) are linear in momenta
so that it appears as linear combination of the creation and annihilation operators.
The last term is cubic in $k$ for which we only need to keep those part with
nonvanishing contribution to the matrix element of Eq.~(\ref{eq:EDSR}).
As we kept the term proportional to $\lambda$ in the original Hamiltonian,
we have to account for the second order terms appearing after the
canonical transformation.
After tedious calculation, we find that
the term proportional to $\gamma\beta$ vanishes while
the term proportional to $\lambda$ remains.
As a result, in the lowest energy level, $n_+=n_-=0$, we obtain from the formula
(\ref{eq:EDSR}) that
\begin{widetext}
\begin{eqnarray}\label{eq:juzhen1}
&& T^{\mathrm  D+R} = -\frac{\lambda}{\hbar
Q_3}\sum_{\nu=+,-}\Bigl\{
  [\omega_c\cos(\varphi-\phi)\cos\theta+i\omega_z\sin(\varphi-\phi)]
  \Omega Q_\nu Q_1
     \nonumber \\
&&+[\Omega\omega_c\cos\theta\sin(\varphi-\phi)-i\omega_z\cos(\varphi-\phi)
(\Omega+\omega_c^2\sin^2\theta)]Q_\nu Q_2\Bigr\} +
\frac{\beta}{\hbar Q_3}\Bigl\{\cos(\varphi-\phi)\times
     \nonumber \\
&&\hspace{3mm} [\Omega\omega_c\cos\theta
 (i\cos2\varphi-\sin2\varphi\cos\theta)
  +\omega_z(\Omega+\omega_c^2\sin^2\theta)
   (\sin 2\varphi-i\cos 2\varphi\cos \theta)]
     \nonumber \\
&& +i\sin(\varphi-\phi)\Omega[\omega_z(i\cos
    2\varphi-\sin 2\varphi\cos\theta)+\omega_c\cos\theta(\sin
     2\varphi-i\cos2\varphi\cos\theta)]\Bigr\}
     -\frac{\gamma}{\hbar Q_3}\Bigl\{
      \nonumber \\
&&\cos(\varphi-\phi)[\Omega\omega_c\cos^2\theta
    +\omega_z(\Omega+ \omega_c^2\sin^2\theta)]
    +i\cos\theta\sin(\varphi-\phi)(\omega_c+\omega_z)\Omega\Bigr\},
\end{eqnarray}
\end{widetext}
where
\begin{eqnarray*}
Q_1  &=&\Bigl(\frac{\cos\theta}{2}+i\cot 2\varphi\Bigr)Q_4
  +(\cos\theta\sin 2\varphi-i\cos 2\varphi) \\
    & & \times(1-\frac{\omega_{+}^2}{\omega_c^2})
    \frac{\omega_c^2}{\Delta^2}~\mathrm {sgn}(\omega_0-\omega_c),
     \\
Q_2&=&\frac{3}{2}iQ_4+(\cos\theta\cos 2\varphi+i\sin 2\varphi)
     \\
&&\times(1-\frac{\omega_{+}^2}{\omega_c^2})
\frac{\omega_c^2}{\Delta^2}~\mathrm {sgn}(\omega_0-\omega_c),
     \\
Q_3&=&\omega_0^2(\omega_c^2\cos^2\theta-\omega_z^2)-\omega_z^2(\omega_c^2-\omega_z^2),
     \\
Q_4&=&-\frac{\omega_c^2}{\Delta^2}\sin 2\varphi\sin^2\theta
 ~\mathrm {sgn}(\omega_0-\omega_c), \\
Q_{\pm} &=& m^{\star}\omega_{\pm}/2\hbar\\
\Omega &=&\omega^2_0 - \omega^2_z.
\end{eqnarray*}
The  terms proportional to  $\beta$ and $\gamma$ in Eq.(~\ref{eq:juzhen1})
are in agreement with the results~\cite{Efros}
for pure Dresselhaus and Rashba spin-orbit couplings.
Then we are in the position to evaluate the EDSR intensity,
$I\propto \mid T\mid^2$, numerically.

\begin{figure}[t]
\includegraphics[angle=-90,width=5.1cm]{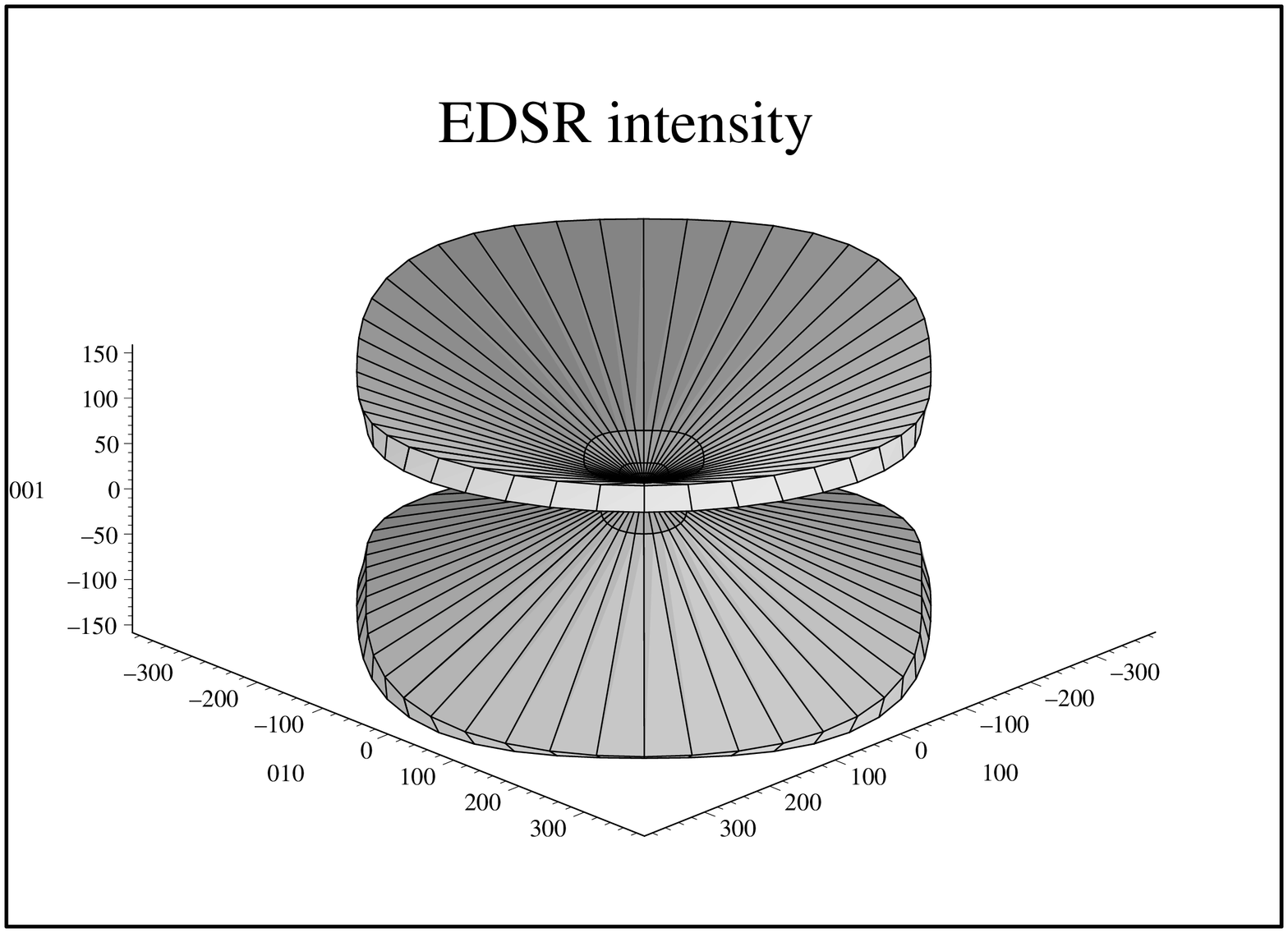}
\includegraphics[angle=-90,width=5.1cm]{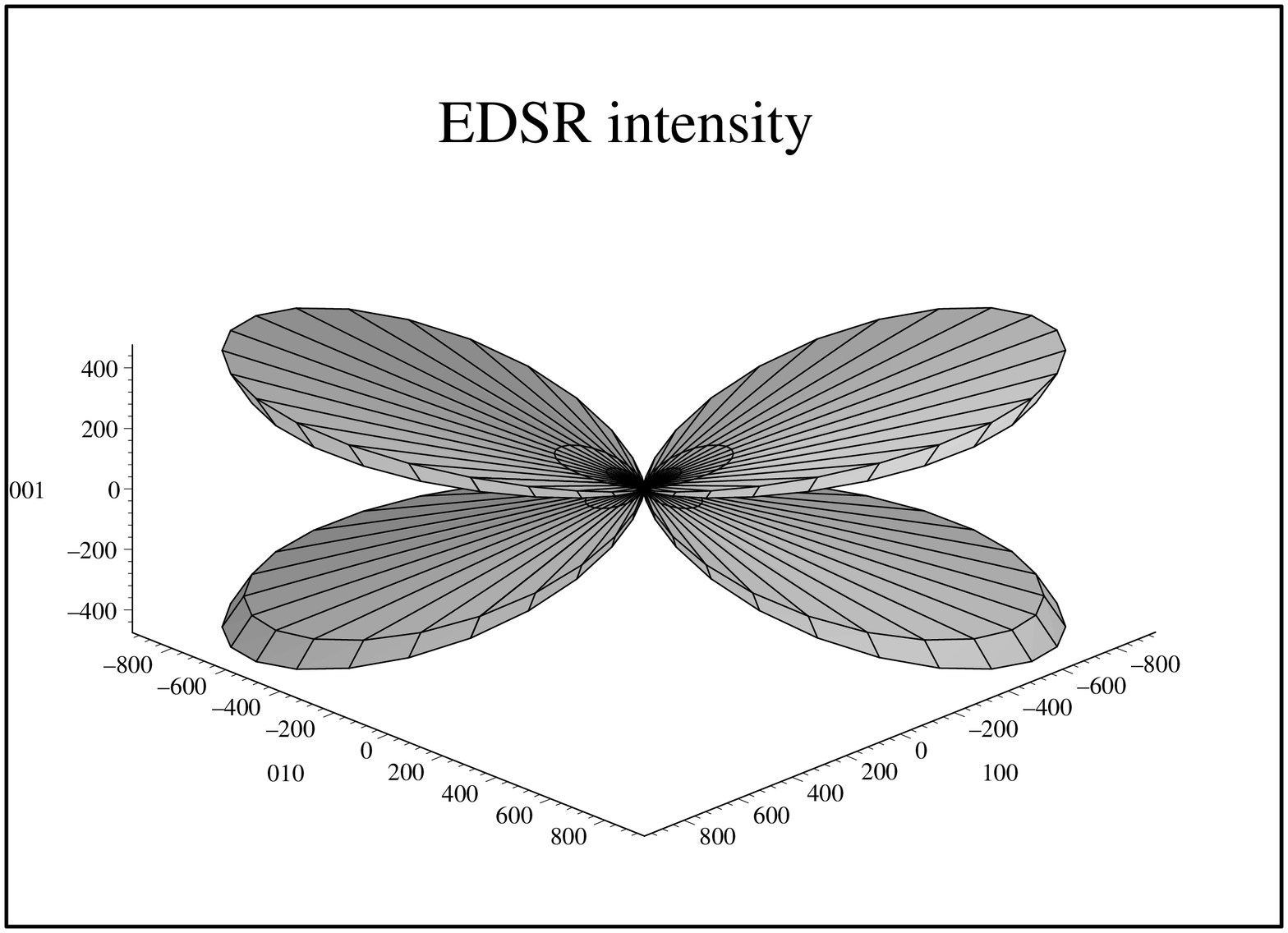}
\caption{\label{fig:direction} The angular dependence of EDSR
intensity $I(\theta, \varphi)$ (arb.units) for a [001] InSb quantum
well with strain tensors
$\epsilon_{xy}=0$ (top panel) and
$\epsilon_{xy}=0.13\%$ (bottom panel). The other parameters are taken as
$\omega_c/\omega_0=0.5$, $\omega_z/\omega_0=-0.16$, $\beta=1$,
$\phi=\pi/4$, an imaginary part $i\delta$($\delta=0.05\omega_0$) was
added to $\omega_c$ to eliminate the divergent pole.}
\end{figure}

Firstly, we investigate the angular dependence of the EDSR intensity
for the InSb based quantum well under strain.
As illustrated in top panel of Fig.~\ref{fig:direction},
the EDSR intensity  will increase when
the magnetic field is tilted. This feature implies that the
manipulation of electron spins  becomes more efficient once a tilted
magnetic field is introduced.
In our calculation, the system's
parameters are chosen by referring to experimental situations,
\ie,
$\lambda=200\, $eV$\AA^3$, $C_3=1.13\times 10^{-7} \mathrm {eVcm}$,
$\omega_z/\omega_c=-0.32$,
$\omega_0=2e B_0 /(m^\star c)$
with $B_0=2\,$T,
$\beta=\lambda\ave{k_z^2}=\lambda m^{\star}\omega_0/(2\hbar)$
and $m^{\star}=0.014m_0$
with $m_0$ being the mass of free electron.
We set $\beta=1$ for convenience in numerical calculation.
It is more important
for us to observe the influence of the strain on the spin
manipulating described by EDSR. The figure in the bottom panel
of Fig.~\ref{fig:direction} shows that the
strain-induced Rashba-type spin splitting makes the characteristic
of the EDSR intensity change significantly. The supremum of the
intensity will increase from $100$ units to over $400$ units after
the shear strain $\epsilon_{xy}=\epsilon_0=0.13\%$,
$\epsilon_0=2\beta/C_3$ is exerted. Additionally, the fourfold
symmetry is broken down to a twofold symmetry.
\begin{figure}[h]
\includegraphics[width=5.9cm]{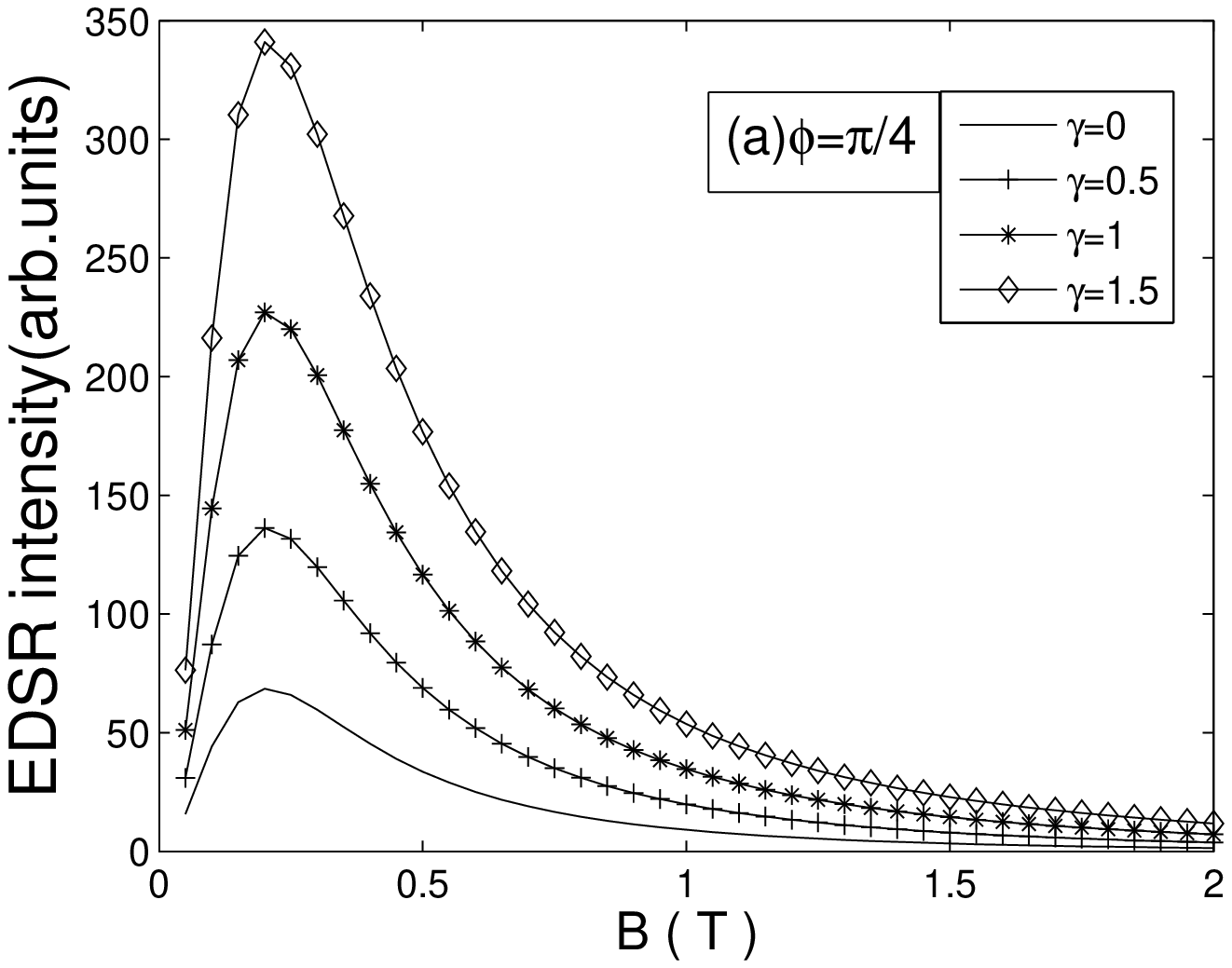}
\includegraphics[width=5.9cm]{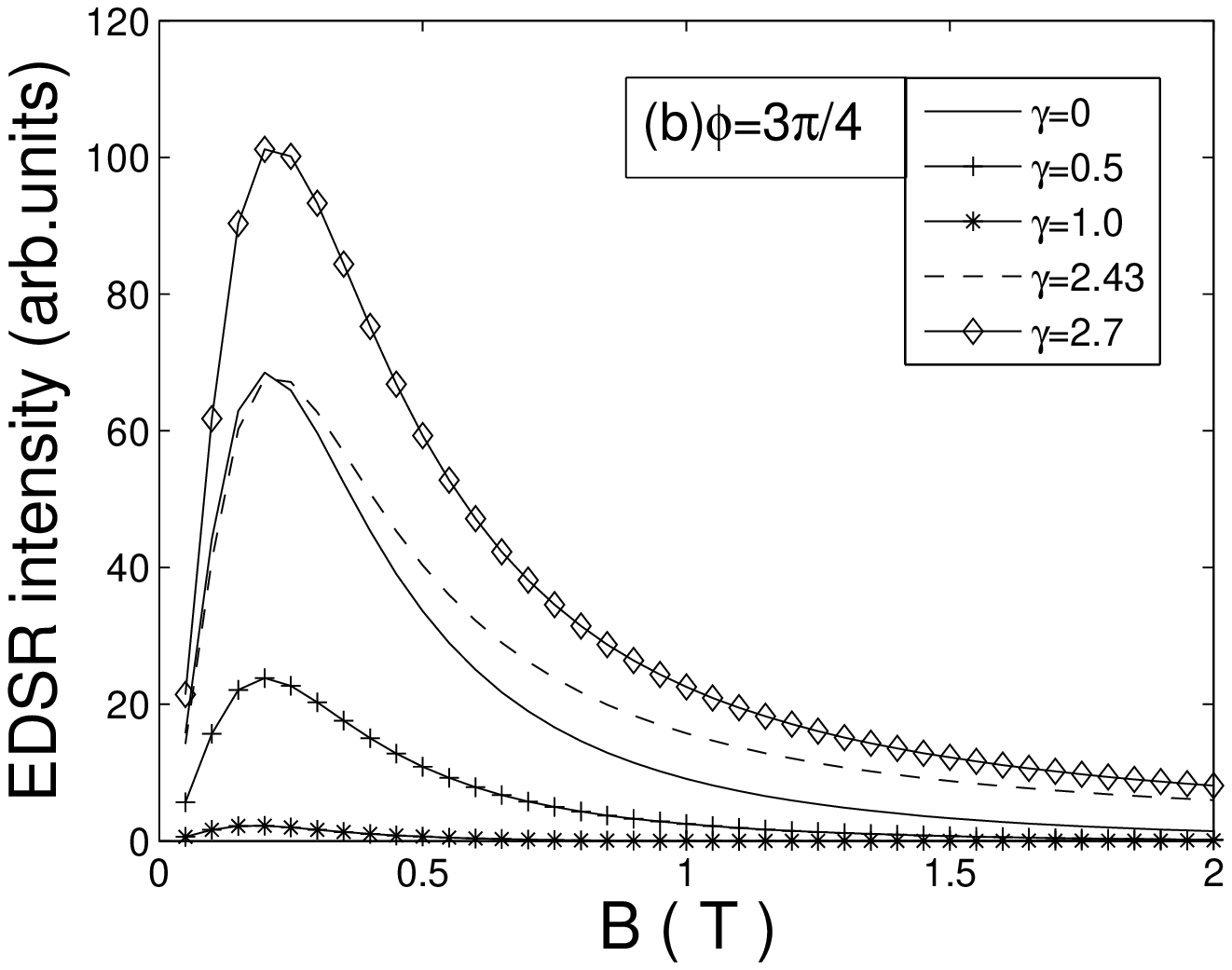}
\caption{\label{fig:Dresselhaus}
The EDSR intensities versus the magnetic field in D+R case for two different
external electric field along the directions
(a) $\phi=\pi/4$ and (b) $\phi=3\pi/4$
with different strain strengths.
Other parameters are taken as
$m^\star = 0.014 m_0$, $\theta=0$, $\varphi=0$,
$\delta=0.05\omega_0$ and $\beta=1$.}
\end{figure}

Furthermore, we analyze the strain-induced effects in the spin manipulation
in different situations. The magnetic field perpendicular to $x$-$y$
plane of the quantum well, saying $\theta=0$ and $\varphi=0$, is
introduced so that our conclusions can be verified by experiments
conveniently.
The dependence of the EDSR intensity on the direction of the
$ac$ electric field is plotted in the appendix,
Fig.~\ref{fig:electricfield} (a), which shows a sinusoid-like behavior
with amplitude and central value being determined by the strain strength.
In Fig.~\ref{fig:Dresselhaus}, we plot
the EDSR intensity versus the magnetic field
for the $ac$ electric field
either along $\phi=\pi/4$ or $\phi=3\pi/4$ directions.
Here, the unit of the EDSR intensity  is $1/\hbar^2\omega_0^2$ with $\omega_0$ being
the same value as in Fig.~\ref{fig:direction};
the other parameters are chosen as
$\beta=1$, $\gamma_0=C_3\epsilon_0/2=1$, and
$\epsilon_{xy}=\gamma \times \epsilon_0$.
As shown in
Fig.~\ref{fig:Dresselhaus}, the EDSR intensities reach an extreme
value at a particular point $B_R$ near $0.2T$ for both cases of
$\phi=\pi/4$ and $\phi=3\pi/4$.
If the confinement frequency $\omega_0$ increases, the magnitude
$B_R$ at which the resonance occurs will increase simultaneously.
Additionally, the location of the resonance peaks will
undergo a slight change if the spin resonance frequency $\omega_z$
differs. Actually, the strain will affect $g$ tensor which is
relevant to $\omega_z$. Thus the strain will bring in a shift in
$B_R$, which is expected to be observed in experiments.

The peak value of EDSR intensity increases from $70$ units to about
$350$ units as the strain strength increases from $\epsilon_{xy}=0$
to $\epsilon_{xy}=0.195\%$ when $\phi=\pi/4$. Four distinct curves
are plotted in Fig~\ref{fig:Dresselhaus}(a), respectively. However,
the variation trends change when the direction of $ac$ electric
field is changed from $\phi=\pi/4$ to $\phi=3\pi/4$ (see
Fig.~\ref{fig:Dresselhaus}(b) ). The peak value of the EDSR
intensity is $70$ units in the strain-free (\ie, $\gamma=0$) case
and decreases if the strain increases. It diminishes to zero when
$\epsilon_{xy}=0.156\%$ (\ie, $\gamma=1.2$), turns to increase after
the strain strength exceeds $\epsilon_{xy}=0.156\%$ and reaches $70$
units again when $\gamma=2.43 $. From Fig.~\ref{fig:Dresselhaus}(b)
together with Fig.~\ref{fig:Dresselhaus}(a), one can see that the
strain-induced Rashba-type spin-orbit coupling combined with
intrinsic Dresselhaus-type coupling can be verified by comparing two
cases for the $ac$ electric field along $\phi=\pi/4$ and
$\phi=3\pi/4$ directions.

Let us make a qualitative analysis with respective to the above
numerical calculations. The contribution of the first term in
Eq.~(\ref{eq:juzhen1}) is relatively small in comparison to the
other terms when the well width is small and/or the temperature is
sufficiently low, thus it can be neglected for a qualitative
estimation. Then diagonal element of spin-flip transition matrix
Eq.~(\ref{eq:juzhen1}) dominates
\begin{eqnarray*}
T^{\mathrm  D+R}\propto &&
-\Omega[(\beta\sin\phi+\gamma\cos\phi)\nonumber
\\
&&-i(\beta\cos\phi+\gamma\sin\phi)].
\end{eqnarray*}
Consequently, the EDSR intensity behaves as
\begin{equation}\label{eq:juzhen11}
I^{\mathrm  D+R}(\phi)\propto \Omega^2(\gamma^2 + 2\gamma\beta\sin 2\phi + \beta^2),
\end{equation}
From this equation, one can see that the EDSR intensity  increases
monotonously when $\phi=\pi/4$ while it firstly decreases and then
increases when $\phi=3\pi/4$ as the strain strength increases.

We also calculated the EDSR intensity when the magnetic field is
tiled ($\theta\neq0$). The corresponding results are similar to that
for $\theta=0$. Therefore electron spins can be sufficiently
manipulated by using an $ac$ electric field for the inversion asymmetry
semiconductor quantum wells exerted with an appropriate strain.

The theoretical result in the above is expected to be measured experimentally.
For this purpose, one may consider such a geometry that the InSb quantum well suffers from
compression along the $[110]$ axis.
The electric field $\mathbf{E}(t)$
is in the plane of the 2DEG, and the magnetic field is perpendicular
to the plane.
A permanent uniaxial stress can be applied to the sample either
with the help of a screw putting in the sample holder
or by means of the other mechanical methods~\cite{Sih}.
In the former approach,
the strength of the uniaxial stress to the sample can be varied
by adjusting the screw before structural failure occurs.
One can measure the EDSR intensities for
the $ac$ electric field along $\phi=\pi/4\,([110])$ and $3\pi/4\,([-110])$
directions under certain strain strength.
The measurement of EDSR intensity  by tuning the strength
of the uniaxial stress can accomplish the verification of theoretical results.
The other approach is due to the fact that strain on samples can be realized
in terms of a technique in growing by molecular beam epitaxy.
A collection of those samples satisfying above strain configuration
with different strain parameters are thus applicable for the same purpose.

\subsection{Strain-induced Dresselhaus-type coupling}

\begin{figure}[t]
\includegraphics[width=5.9cm]{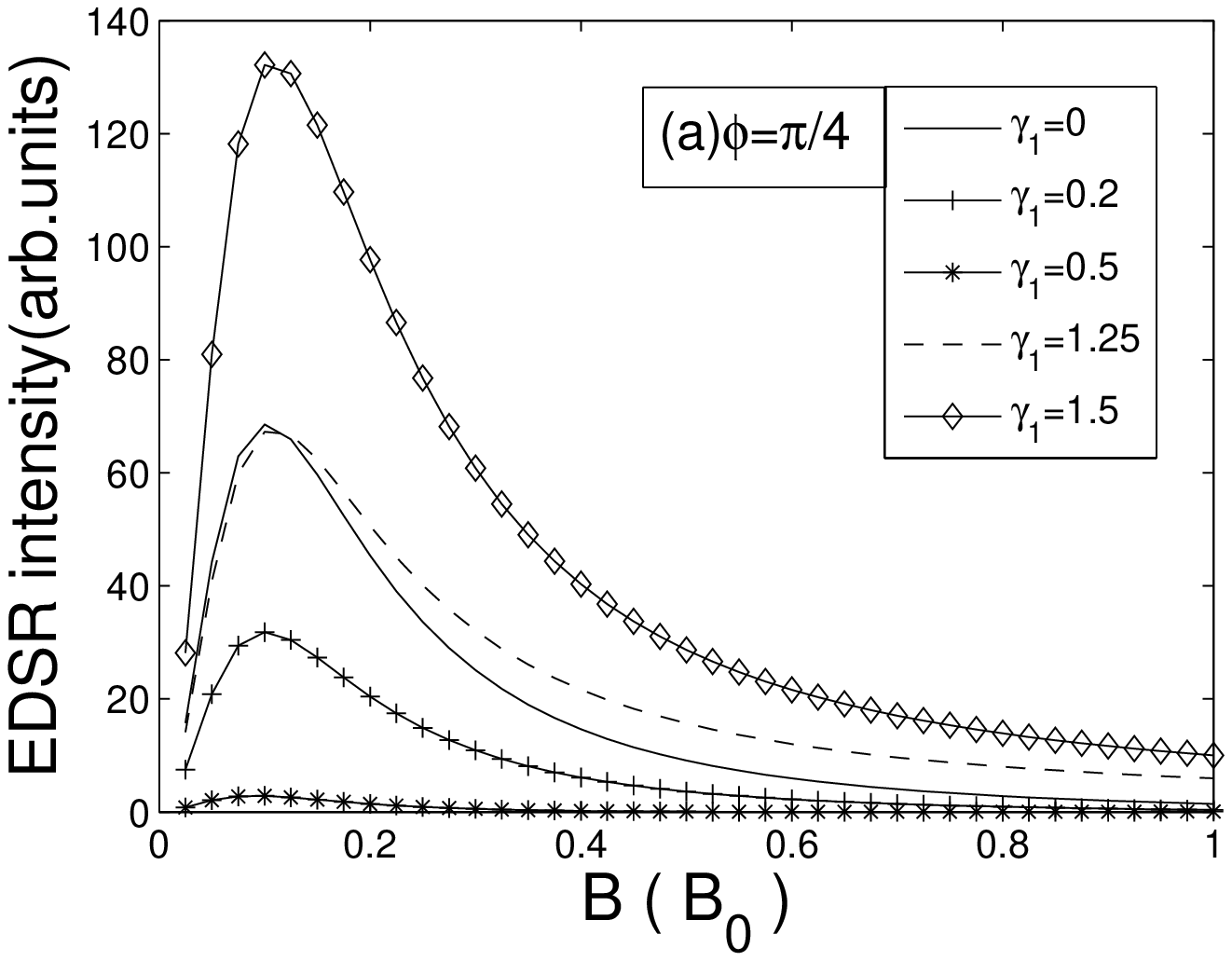}
\includegraphics[width=5.9cm]{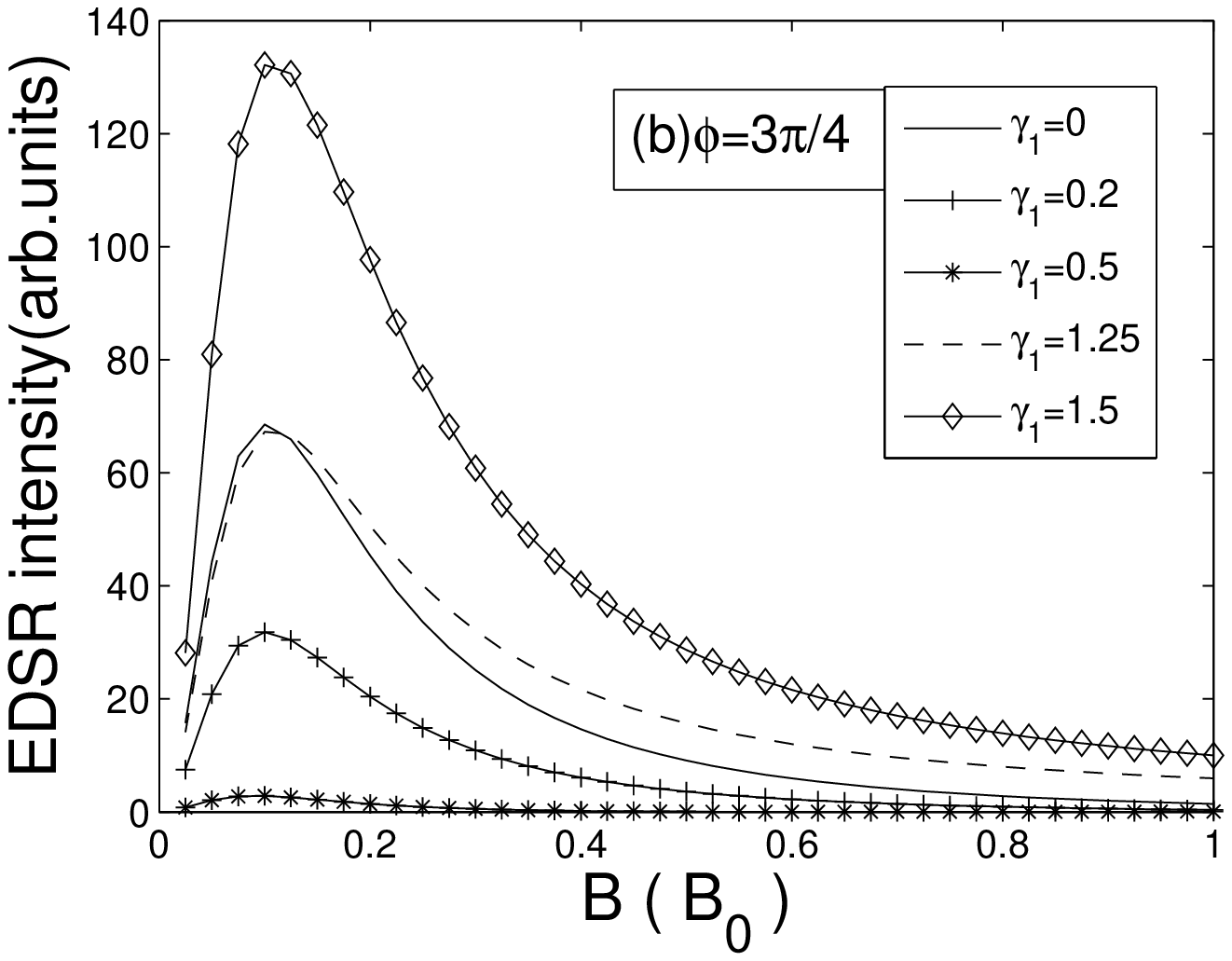}
\caption{\label{fig:DrDesselhaus}
The EDSR intensities versus the magnetic field in D+D case are plotted for
(a) $\phi=\pi/4$ and (b) $\phi=3\pi/4$.
Here $B_0= \omega_0m^\star c/2e$ and the other parameters are taken as
$\theta=0$, $\varphi=0$, $\delta=0.05\omega_0$, $\beta=1$,
$\gamma_{10}=\beta$
and $\omega_z/\omega_c=-0.32$.}
\end{figure}

As a comparison, we consider other kind of semiconductor
quantum wells.
If the spin splitting induced by strain is assumed to
be Dresselhuas-type, saying $H_{st}^{\mathrm  D}$, while $H_{in}^{\mathrm  D}$ is still
dominant over $H_{in}^{\mathrm  R}$ in these quantum wells, the total
Hamiltonian describing the spin precession for electrons in such
quantum wells (call D+D case for brevity) is written as
\begin{eqnarray}
H_{so}^{\mathrm  D+D} &=& H_{in}^{\mathrm  D}+H_{st}^{\mathrm  D}
        \nonumber \\
      &=&\sigma_xk_x(\lambda k_y^2-\beta)+\sigma_yk_y(\beta-\lambda k_x^2)
        \nonumber \\
     & & +\gamma_1(\sigma_xk_x-\sigma_yk_y),
\end{eqnarray}
with the strain configuration $\epsilon_{xx}=\epsilon_{yy}$ and
$\gamma_1=D(\epsilon_{zz}-\epsilon_{xx})>0$. The parameters
$\lambda$, $\beta$ and $\gamma_1$ are different for different kind of
quantum wells.
In the following part of this paper, we do not consider concrete materials
in numerical calculation.
In the lowest energy level $n_+=n_-=0$, the diagonal element of the
spin-flip transition matrix $T^{\mathrm D+D}$ for $H_{so}^{\mathrm  D+D}$
is calculated  by using the similar method we
employed in previous subsection,
\begin{widetext}
\begin{eqnarray}
T^{\mathrm  D+D} &=&-\frac{\lambda}{\hbar Q_3}\sum_{\nu=+,-}
 \Bigl\{\bigl[
  \omega_c\cos(\varphi-\phi)\cos\theta+i\omega_z\sin(\varphi-\phi)
  \bigr]\Omega Q_\nu Q_1
   \nonumber \\
&& +\bigl[\Omega\omega_c\cos\theta\sin(\varphi-\phi)
  -i\omega_z\cos(\varphi-\phi)
   (\Omega+ \omega_c^2\sin^2\theta)
   \bigr]Q_\nu Q_2
 \Bigr\}
+\frac{(\beta-\gamma_1)}{\hbar Q_3}\Bigl\{
\nonumber \\&&
  \cos(\varphi-\phi)\bigl[\Omega\omega_c\cos\theta
 (i\cos2\varphi-\sin2\varphi\cos\theta)
+\omega_z(\Omega+\omega_c^2\sin^2\theta)
 (\sin 2\varphi-i\cos 2\varphi\cos \theta)\bigr]
  \nonumber \\
&&+i\sin(\varphi-\phi)\Omega \bigl[\omega_z(i\cos2\varphi-\sin
2\varphi\cos\theta)
 +\omega_c\cos\theta(\sin 2\varphi-i\cos 2\varphi\cos\theta)
\bigr]\Bigr\}.
\end{eqnarray}
\end{widetext}
We plot the EDSR intensity versus the magnetic field
in the unit of $B_0= \omega_0m^\star c/2e$
for the D+D case in Fig.~\ref{fig:DrDesselhaus}.
We have chosen $\omega_z/\omega_c = -0.32$ and
set $\beta=1$, $\gamma_{10}=\beta$,
and $\epsilon_{zz}-\epsilon_{xx}=\gamma_1\times\gamma_{10}/D$ for convenience.
Although the ratio $\omega_z/\omega_c$ may differ for different
quantum wells, the feature of the curves in Fig.~\ref{fig:DrDesselhaus}
can manifest the main characteristics of the
EDSR intensity for this kind of spin-orbit coupling.
Since the concrete material is not specified here,
we can only investigate the variation trends of the EDSR intensity
with different strain strengths, which is adequate for the
comparison between different combinations of spin-orbit couplings.
As the strain strength increases, seen from
Fig.~\ref{fig:DrDesselhaus}, the EDSR intensities decrease at first;
after the strain strength exceeds certain value it turns to increase
in both cases $\phi=\pi/4$ and $\phi=3\pi/4$.
Actually, the EDSR intensity is independent of the direction of
the $ac$ electric field (see Fig.~\ref{fig:electricfield}(b)).
The distinct features in Fig.~\ref{fig:Dresselhaus} and
Fig.~\ref{fig:DrDesselhaus} are therefore suggested to identify whether the
strain-induced spin splitting in a quantum well is Dresselhuas-type
or Rashba-type by measuring the EDSR intensity in experiments.
\begin{figure}[t]
\includegraphics[width=5.9cm]{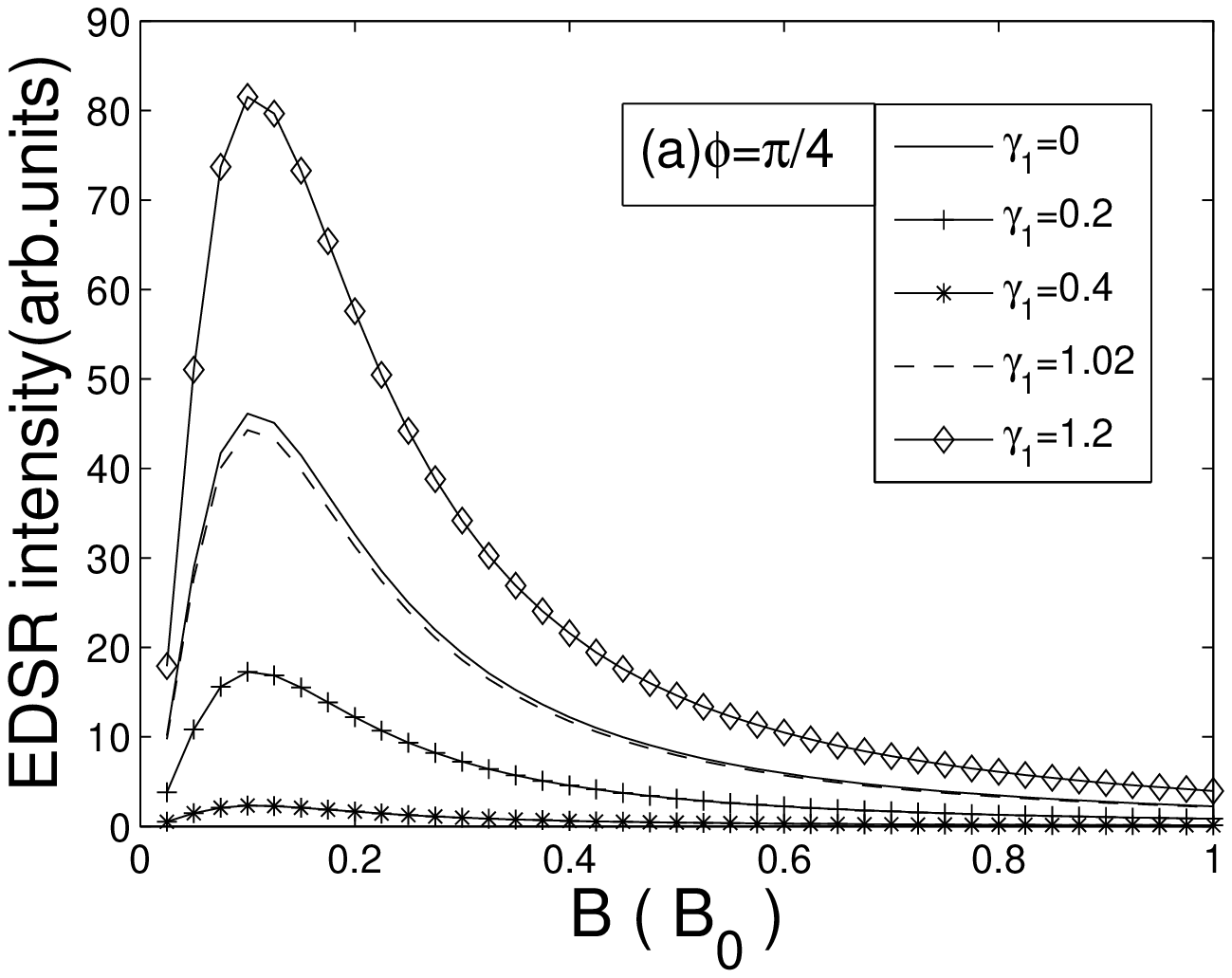}
\includegraphics[width=5.9cm]{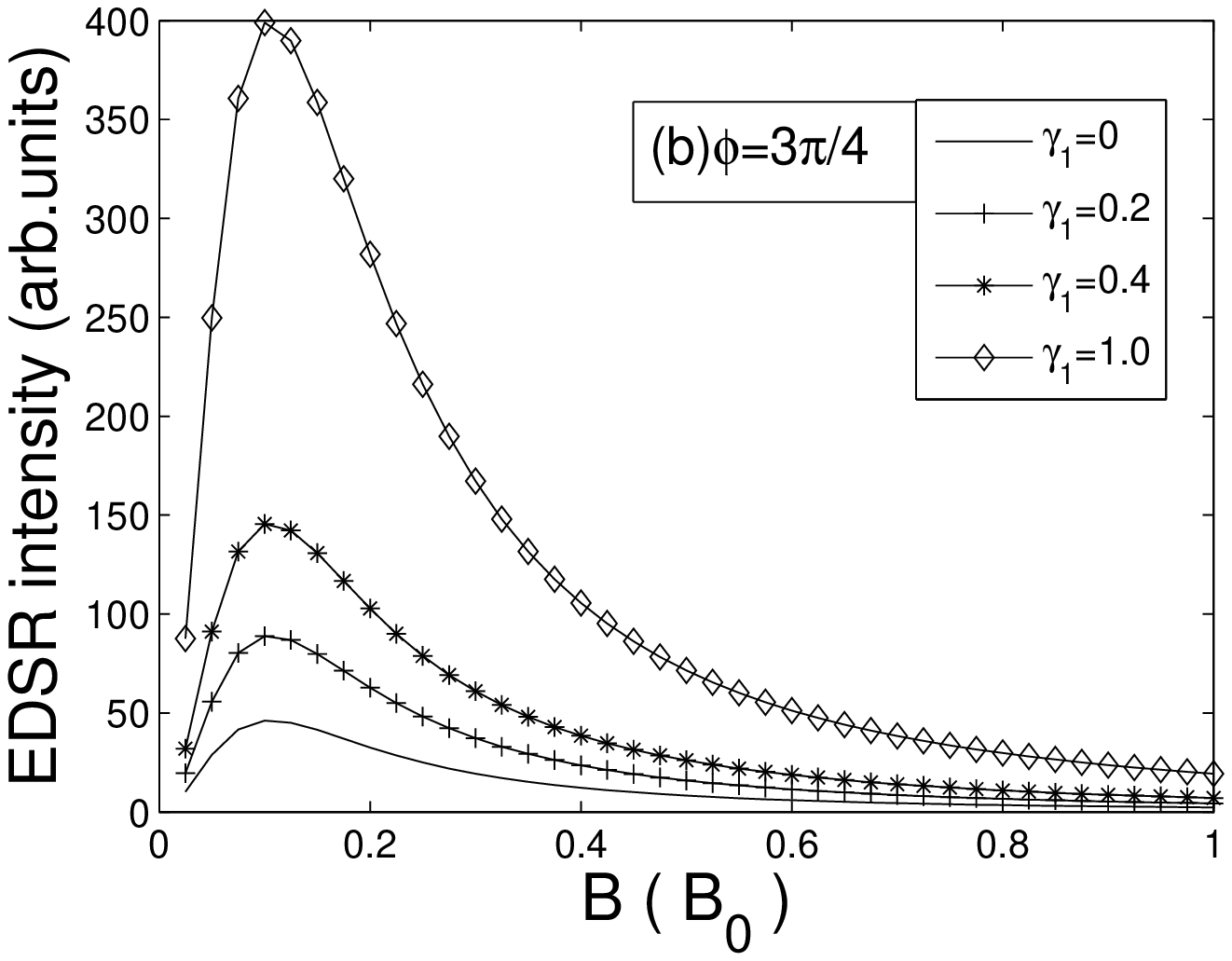}
\caption{\label{fig:Rashba}
The EDSR intensities versus the magnetic field in R+D case
are plotted for (a) $\phi=\pi/4$ and (b) $\phi=3\pi/4$.
Here $B_0= \omega_0m^\star c/2e$ and the other parameters
are taken as $\theta=0$, $\varphi=0$,
$\delta=0.05\omega_0$, $\alpha=1$, $\gamma'_{10}=\alpha$
and $\omega_z/\omega_c=-0.32$.}
\end{figure}

\section{Systems with intrinsic Rashba coupling}\label{sec:Rashba}

In this section, we consider the characteristics of EDSR for quantum
wells with intrinsic Rashba-type spin splitting due to structure inversion asymmetry,
which is described by the Hamiltonian $H_{in}^{\mathrm  R}$.
Since the strain can introduce spin splittings of either Rashba-type
$H_{st}^{\mathrm  R}$ or Dresselhaus-type $H_{st}^{\mathrm  D}$,
the total Hamiltonian corresponding to R+R and R+D cases are, respectively,
\begin{eqnarray}
H_{so}^{\mathrm  R+R} &=&
\alpha(\sigma_xk_y-\sigma_yk_x)+\gamma(\sigma_xk_y-\sigma_yk_x),
   \nonumber\\
H_{so}^{\mathrm  R+D} &=&
\alpha(\sigma_xk_y-\sigma_yk_x)+\gamma_1(\sigma_xk_x-\sigma_yk_y).
   \label{eq:hami1-2}
\end{eqnarray}
Here the definitions of parameters $\alpha$, $\gamma$ and $\gamma_1$ are the same
as in the above, while their magnitudes depend on concrete materials and strain
configurations.
The Hamiltonians in Eqs. (\ref{eq:hami1-2})
describe systems with Rashba-type and Dresselhaus-type
strain-induced spin splittings, respectively.
Their corresponding diagonal elements of spin-flip transition matrices
$T^{\mathrm R+R}$ and $T^{\mathrm R+D}$ are respectively obtained,
\begin{eqnarray}\label{eq:juzhen3}
T^{\mathrm R+R}&=& -\frac{(\gamma+\alpha)}{\hbar Q_3}
\Bigl\{\cos(\varphi-\phi)\times 
  \nonumber\\
&&\bigl[\Omega\omega_c\cos^2\theta
 +\omega_z(\Omega+\omega_c^2\sin^2\theta)\bigr]
  \nonumber\\
&&+i\cos\theta\sin(\varphi-\phi)(\omega_c+\omega_z)\Omega
 \Bigr\},
\end{eqnarray}
\begin{eqnarray}\label{eq:juzhen4}
T^{\mathrm  R+D}
= \displaystyle\frac{-\gamma_1}{\hbar Q_3}\Bigl\{
  \cos(\varphi-\phi)\times \hspace{34mm}
   \nonumber\\
 \bigl[\Omega\omega_c\cos\theta(i\cos2\varphi-\sin2\varphi\cos\theta)
   \nonumber \\
+\omega_z(\Omega+\omega_c^2\sin^2\theta)
     (\sin 2\varphi-i\cos2\varphi\cos \theta)\bigr]
     \nonumber \\
+i\sin(\varphi-\phi)\Omega
  \bigl[\omega_z(i\cos2\varphi-\sin 2\varphi\cos\theta)
  \nonumber\\
 +\omega_c\cos\theta(\sin 2\varphi-i\cos 2\varphi\cos\theta)\bigr]
\Bigr\}
   \nonumber \\
-\frac{\alpha}{\hbar Q_3}\Bigl\{\cos(\varphi-\phi)
\bigl[\Omega\omega_c\cos^2\theta
 +\omega_z(\Omega+ \omega_c^2\sin^2\theta)\bigr]
 \nonumber\\
 +i\cos\theta\sin(\varphi-\phi)(\omega_c+\omega_z)\Omega\Bigr\}.
\end{eqnarray}

For the R+R case, one can see from Eq.~(\ref{eq:juzhen3})
that the EDSR intensity increases monotonously
as the strain strength increases since
$T^{\mathrm R+R}\propto(\gamma+\alpha)$ for either $\phi=\pi/4$ or
$\phi=3\pi/4$.
For R+D case,
using $T^{\mathrm R+D}$ in Eq.~(\ref{eq:juzhen4}),
we plot the EDSR intensity versus the magnetic field
in the unit of $B_0= \omega_0m^\star c/2e$
in Fig.~\ref{fig:Rashba}.
The parameter choices of the plot are $\omega_z/\omega_c=-0.32$,
$\theta=0$, $\varphi=0$, $\delta=0.05\omega_0$, $\alpha=1$,
$\gamma'_{10}=\alpha$ and
$\epsilon_{zz}-\epsilon_{xx}=\gamma_1\times \gamma'_{10}/D$.
Clearly, the intensity firstly decreases to zero and then increases after the
strain strength exceeds $\gamma_1=0.5$ for $\phi=\pi/4$ (see
Fig.~\ref{fig:Rashba}(a) ), which differs from the case in
Fig.~\ref{fig:Dresselhaus}(a). As seen from Fig.~\ref{fig:Rashba}(b)
that the intensity increases monotonously along with the increment
of strain strength when $\phi=3\pi/4$, which is also different from
the case in Fig.~\ref{fig:Dresselhaus}(b).
The aforementioned features are consistent with
Figs.~\ref{fig:electricfield}(c) and \ref{fig:electricfield}(d)
where the dependences of EDSR intensity on the direction
of $ac$ electric field are illustrated for
R+R and R+D cases, respectively.

\section{Conclusion}

We have analyzed the strain-assisted manipulation of electron spins
in quantum wells where an $ac$ electric field in $x$-$y$ plane and a
perpendicular magnetic field are applied. The strain effects on the
manipulation are different for different semiconductor quantum
wells. We exhibited that the efficiency of electron-spin
manipulation can be enhanced by tuning the strain strength
of the sample. There are four compositions for intrinsic and
strain-induced spin splittings,
namely, D+R, D+D, R+R, or  R+D.
These four situations can be distinguished from each
other by changing the direction of the $ac$ electric field from
$\phi=\pi/4$ to $\phi=3\pi/4$ and tuning the strain strengths in
EDSR experiments. The effect combining those four kinds of spin
splittings is helpful for understanding the experimental phenomena
brought in by strain in some semiconductor quantum wells.

The work was supported by Program for Changjiang Scholars and Innovative
Research Team in University, and by NSFC grant Nos. 10225419 and 10674117.

\appendix\section{}
The dependence of EDSR intensities on the direction of the applied electric field
are plotted for various strain strengths:
\begin{figure}[t]
\hspace{-1.4cm}\includegraphics[width=10.0cm]{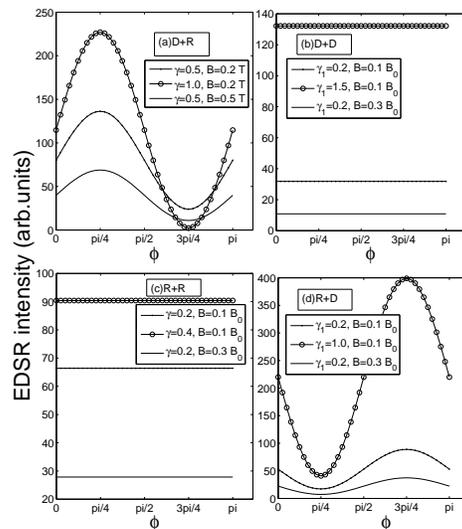}
\vspace{-0.6cm}
\caption{\label{fig:electricfield}
The EDSR intensities versus $\phi$,
the direction of the $ac$ electric field,
with parameters
$\omega_z/\omega_c=-0.32$, $\theta=0$ and $\varphi=0$ for the four cases:
(a) D+R case, $\beta=1$, $\epsilon_{xy}=\gamma\times \epsilon_0$,
$\epsilon_0=2\beta/C_3=0.13\%$, $\omega_0=2eB_0/(m^{\star}c)$ with
$B_0=2\,$T.
(b) D+D case, $\omega_0=2eB_0/(m^{\star}c)$,
$\epsilon_{zz}-\epsilon_{xx}=\gamma_1\times\gamma_{10}/D$,
$\beta=1$ and $\gamma_{10}=\beta$;
(c) R+R case,
$\omega_0=2eB_0/(m^{\star}c)$, $\epsilon_{xy}=\gamma\times
\epsilon_0'$, $\epsilon_0'=2\alpha/C_3$ and $\alpha=1$;
(d) R+D case, $\omega_0=2eB_0/(m^{\star}c)$,
$\epsilon_{zz}-\epsilon_{xx}=\gamma_1\times\gamma_{10}'/D$,
$\alpha=1$ and $\gamma_{10}'=\alpha$.}
\end{figure}

\end{document}